\documentstyle[11pt,paspconf,epsf]{article}

\begin{document}

\title{Stabilization of Large Scale Structure \\by Adhesive
Gravitational Clustering}

\author{T. Buchert\altaffilmark{1}}
\affil{Theory Division, CERN, CH--1211 Geneva 23, Switzerland}

\altaffiltext{1}{on leave from: Theoretische Physik, 
Ludwig--Maximilians--Universit\"at, Theresienstr. 37, D--80333 M\"unchen, 
Germany; email: buchert@theorie.physik.uni-muenchen.de}

\begin{abstract}
The interplay between gravitational and dispersive forces in a multi--streamed
medium leads to an effect which is exposed in the present note as
the genuine driving force of stabilization of large--scale structure.
The conception of `adhesive gravitational clustering' is advanced to 
interlock the fairly well--understood epoch of formation of large--scale
structure and the onset of virialization into objects that are dynamically
in equilibrium with their large--scale structure environment.
The classical `adhesion model' is opposed to a class of more general
models traced from the physical origin of adhesion in kinetic theory. 
\end{abstract}



\section{Previrialization and Adhesion.}

`Adhesive gravitational clustering' touches on the basis of why structures
stabilize after their formation. On the one hand, analytical understanding of structure
{\em formation} is well advanced (see, e.g., the review by Sahni \& Coles 1995). 
However, models that evolve inhomogeneities
into the nonlinear regime also predict structure decay after their formation. 
On the other hand, the understanding of {\em virialization} is hosted in 
stellar systems theory of classically isolated bound objects. What 
are the equilibrium structures that could be called `relaxed' while 
still interacting gravitationally with their large--scale structure environment ?
The fact that the transition to `virialized' systems is not immediate is mirrored in
expressions such as `previrialization', invoked by Peebles and 
collaborators (Davis \& Peebles 1977, Peebles 1990, Lokas et al. 1996).
The `adhesion approximation' as invented by Gurbatov et al. (1989) 
takes its title from the phenomenological
observation that structures `stick together' after shells of cold matter
cross; numerical simulations predict multiple shell--crossings due to
dragging forces that prevent the particles from escaping high--density
regions. We shall identify these forces in the framework of kinetic theory.
It is here where an important period of `adhesive clustering' 
sets in, which eventually leads to the type of bound objects that we
observe in the Universe and that may finally populate {\em fundamental planes}
in spaces spanned by integral properties of these objects
(Fritsch \& Buchert 1999). 

\section{The formation of large--scale structure.}

\begin{figure}
\epsfxsize=12.0cm
\epsffile{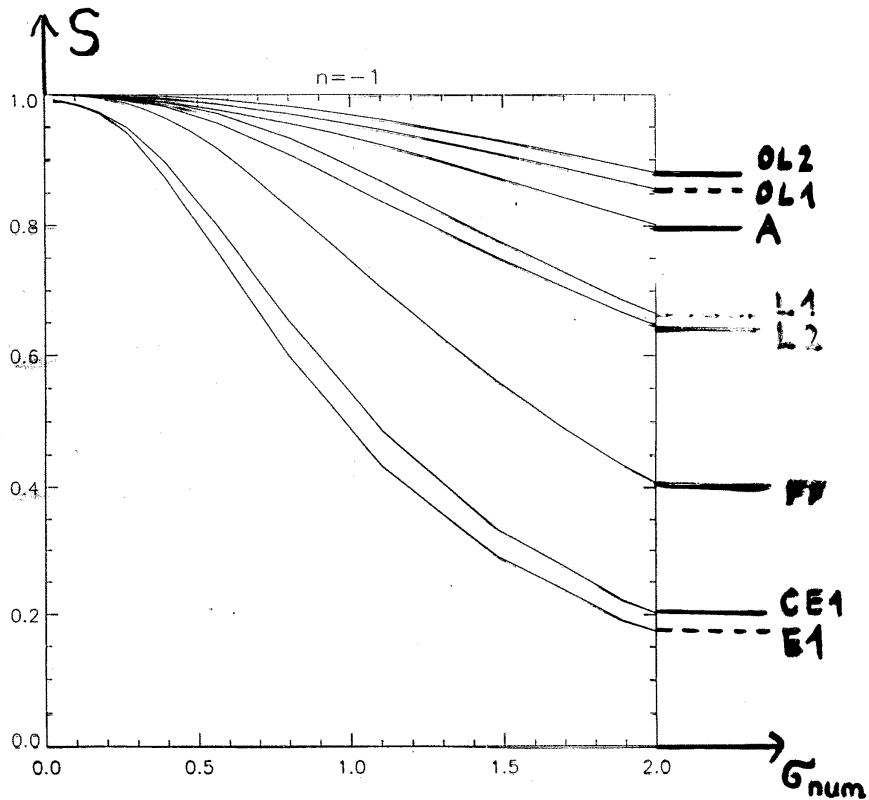}
\caption{Comparison of analytical approximation schemes:
Displayed is the cross--correlation coefficient 
$S=<\frac{\delta_{\rm num}\delta_{\rm approx}}{\sigma_{\rm
num}\sigma_{\rm approx}}>$ that compairs the density contrast fields $\delta$ of a
variety of analytical schemes (normalized to their r.m.s. values $\sigma$) and 
averaged over some scale as a function
of wavenumber; from top to bottom:
Optimized Lagrangian approximations (second--order (Melott et al. 1995)
and first--order (Coles et al. 1993, Melott et al. 1994)),
Adhesion model (Gurbatov et al. 1989, Weinberg \& Gunn 1990), 
Zel'dovich approximation (Zel'dovich 1970), Second--order Lagrangian
approximation (Buchert \& Ehlers 1993), 
Frozen--Potential approximation (Bagla \& Padmanabhan 1994, Brainerd et al. 1994), 
Linear theory (`chopped', i.e. with adapted average, and
unchopped).
S measures whether mass is moved to the right place, S = 1 means
ideal congruence of both fields.} 
\label{fig-1}
\end{figure}

The understanding of large--scale structure formation may be centred on
Zel'do\-vich's idea: just move the particles under inertia and
follow them until they cross. For suitably scaled variables the 
exact solution of a force--free continuum reproduces the linear theory
of gravitational instability in the linear limit (Zel'dovich 1970, 
Zel'dovich \& Myshkis 1973, Shandarin
\& Zel'dovich 1989, Buchert 1989). 

Taking Zel'dovich's simple view at face value, gravity is not important at all
to form structures like curved sheets, filaments and clusters, their 
formation is merely 
an effect of focussing of trajectories we are familiar with in geometrical
optics of light rays. The resulting network of {\it caustics} is unstable:
just increase the depth (advance the time) of a water basin (the Universe)
on the bottom of which (at the time of shell--crossing)
you see a network of high--intensity (high--density) structures, 
and you have this phenomenon. Indeed, Zel'dovich's model applied to 
generic initial data, say powerlaw spectra with slopes in the range 
$n < -3$ at the high--frequency end, reproduces astonishingly well the
density fields predicted by N--body runs of the same initial data
(Melott et al. 1995). 
`Pancake models', which don't have structure on small scales in the
initial conditions, fall in this category (Buchert et al. 1994).   
The spectral index $n=-3$ distinguishes between two different scenarios
of structure formation: either structures form starting from large scales
(top--down), or they form starting hierarchically from small scales
(bottom--up); the total power in some wave number interval has logarithmic
dependence in the case $n=-3$ implying that structures on every scale form
at about the same instant, since all wave numbers were given similar initial
amplitudes. In the other cases $n > -3, n < -3$ the integrated power 
has powerlaw shape; hence, $n=-3$ defines clearly distinct regimes.

Truncation of high--frequency information in the sense of cutting
down the amplitude of short--wavelength perturbations 
to the `critical spectrum' $n=-3$ results in 
optimized schemes which can compete with N--body runs down to the scales
of galaxy clusters 
(Coles et al. 1993, Melott et al. 1994). Refining Zel'dovich's step is 
possible in the framework of the Lagrangian perturbation theory, 
where it can be re--evaluated within the full gravitational context 
as a first--order solution (Buchert 1989; 1996 and
ref. therein).  As a consequence the performance can be improved
(Melott et al. 1995); `optimized Lagrangian schemes' work even for 
CDM--type intial data (Wei{\ss} et al. 1996).
Also, other characteristics of the matter distribution are matching the 
expected (Bouchet et al. 1995).
 
Figure~\ref{fig-1} shows a comparison of N--body density fields with
those predicted by various analytical schemes (some of them will be 
described below): the `adhesion
approximation' works in the right direction and shows better performance
than the Lagrangian schemes. Still, if the Lagrangian schemes are 
optimized, the `adhesion approximation' falls short indicating that the
model is not satisfactorily tailored for the gravitational multi--stream effect
it was launched to describe. 
The spectral index of a powerlaw spectrum was $n=-1$ which is an extreme
test of the approximations: much power on small scales even results in the 
superiority of the first--order over the second--order Lagrangian approximation,
since the latter accelerates structure decay after caustics have formed.

In the following we shall encounter a broader
range of possible models that show how to consolidate and generalize
the adhesion idea.

\newpage

\begin{figure}
\epsfxsize=11.0cm
\epsffile{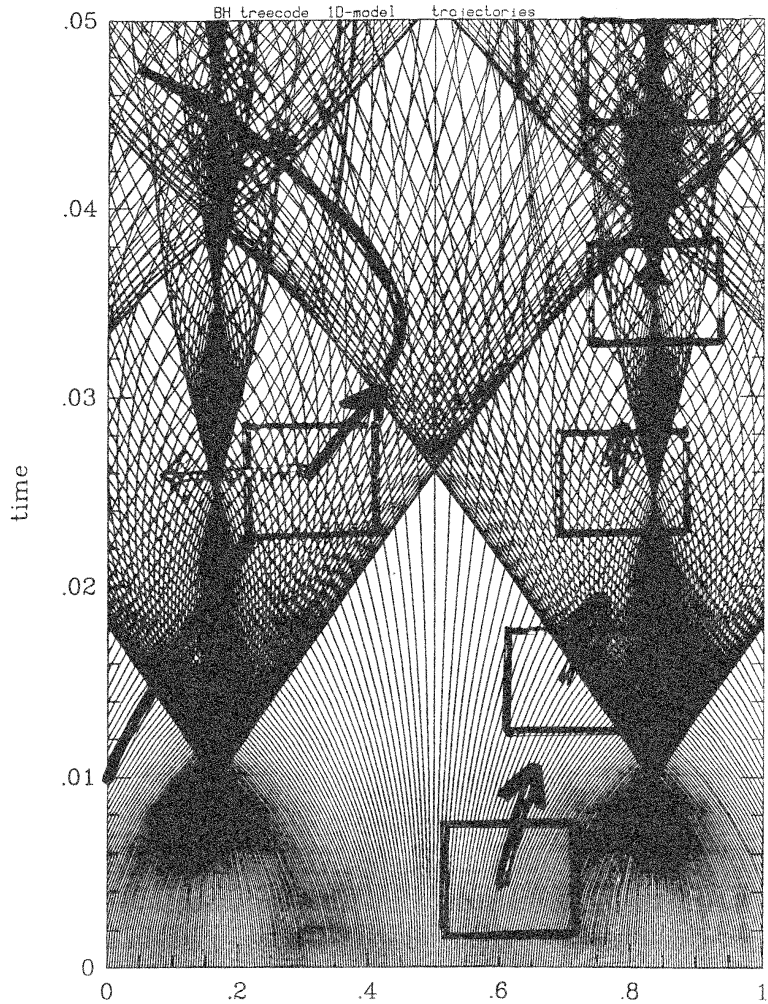}
\caption{The coarse--graining idea is exemplified for a family of crossing
trajectories from a 2D tree--code simulation: we appreciate
the oscillatory behavior of innermost trajectories that are kept back
from escaping the high--density region due to multi--stream forces;
within the multi--stream regions there are many streams (velocities) at a 
given Eulerian position. 
Two possible situations are highlighted for a coarse--grained volume element 
with oscillatory bulk velocity (left) and smoothly decaying bulk velocity (right): 
the kinetic energy of the bulk motion
is transformed into internal kinetic energy of the coarse--grained element
as the element moves into the beast.} 
\label{fig-2}
\end{figure}

\section{Adhesion: effective dynamics of multi--stream systems.}

Working towards a model that continues to be valid after 
shell--crossing singularities develop may be grounded on 
multi--stream hydrodynamics of interpenetrating streams of a cold
medium. This way to go, although being logically the most 
adequate one, may not have the broad range of expression that we 
gain by moving to a kinetic description. The conception to be presented
below, however, is a phenomenological one. We keep the notion of 
a continuous (coarse--grained) fluid element, but we consider its motion
in phase space. Thus we are dealing with 
a one--particle distribution function, or simply a phase space density,
neglecting terms that would arise  by coarse--graining the
N--particle distribution function.  The basic system of equations therefore
consists of the so--called Vlasov--Poisson system. By averaging out the
velocity information we end up with a set of equations that describe an
effective dynamics of a possibly multi--streamed medium.
(For details see: Buchert \& Dom\'\i nguez 1998). This effective dynamics will
in general entail a hierarchy of evolution equations for the velocity 
moments of the phase space density. Here is what I shall call the 
``phenomenological closure'' condition: imagine the shape of a fluid element in 
velocity space that crosses a caustic in real space. The so--called 
velocity ellipsoid, centered on a fluid element with a bulk 
velocity that originally coincides with the velocities of individual particles, 
will expand in a highly anisotropic fashion; as soon as a three--stream 
system develops, two of the streams will head off in almost opposite 
directions and the third (decaying) stream will have small velocity, so that 
overall the ellipsoid starts out to be a flattened disk. Only later, when more
and more streams develop, the ellipsoid has a chance to become spherical
(``virialized''). We are going to mask this ellipsoid with a sphere for all times
and look at the effective multi--stream force that is excerted on the bulk
motion ${\bar {\bf v}} = \langle {\bf v} \rangle$ of the coarse--grained 
fluid element. Mathematically, we require that the 
velocity dispersion tensor be idealized by an isotropic tensor:

\begin{equation}
\Pi_{ij} = \varrho (\langle {\bf v}{\bf v}\rangle - 
{\bar {\bf v}}{\bar {\bf v}}) = 
p \delta_{ij}\;\;\;\;;\;\;\;\;p = \alpha (\varrho)\;>\;0\;\;\;.
\label{pressure} 
\end{equation}
The last equation requires in addition that the multi--stream pressure 
should be given as a function of the density only. This is the `poor man's way'
to close the hierarchy. 

\noindent
In cosmology we study 
the evolution equations for the two dynamical fields peculiar--velocity
$\bf u$ and peculiar--acceleration $\bf w$, or the density $\varrho$, 
respectively. The corresponding effective
evolution equations for these fields with the assumptions sketched above are
$(H:=\frac{\dot a}{a})$:
\begin{equation}
\partial _{t} \varrho + 3 H \varrho + {1 \over a} (\varrho u_{i})_{,i} = 0\;\;\;\;\;\; , 
\;\;\;\;\;\; \partial_{t} {u}_{i} + {1 \over a} {u}_{j} u_{i,j} + H {u}_{i} 
= w_{i} -{\alpha' \over \varrho a} \varrho_{,i} \;\;\;\;\;\;,
\label{evolution}
\end{equation}
subjected to the Newtonian field equations
$w_{i,i}=-4 \pi G a (\varrho - \varrho_H), w_{i,j} = w_{j,i}.$
Note that $\bf u$ is the bulk velocity of a fluid element in coordinates that 
are comoving with the Hubble flow (the overbar is omitted for simplicity,
and a comma means spatial derivative with 
respect to these coordinates). Figure 2 illustrates
how this bulk flow will give away its kinetic energy into internal kinetic 
energy of the fluid element; the elements acquire a ``temperature'' when
they move into the multi--stream region, while the amplitude of the 
bulk flow decays.

\subsection{Old Adhesion.}

One of the most committed of all approaches to multi--streaming
is furnished by Burgers' equation. An exact solution in 3D is available.
Still, numerical techniques await a highly resolved insight into the
multi--stream regime. Instead, the ``burgerlencing'' effect,
being quintessential for holding
fluid elements together inside high--density peaks, is roughhewned into the
phenomenological language of `adhesion'. 
A smart way to employ Burgers' equation on the cosmological stage was
proposed by Gurbatov et al. (loc.cit.). Below, we give their formal arguments
leading to the `adhesion approximation', which we right after that shall derive 
from kinetic theory following Buchert \& Dom\'\i nguez (loc.cit.).
  
Certainly one  of the simplest ways of stating the formal structure of
the `Zel'dovich approximation' (Zel'dovich
1970, 1973) is to postulate a law of motion of the form
\begin{equation}
{\bf w} = F(t) {\bf u}\;\;\;.
\label{proportionality}
\end{equation}
If gravity drags into the direction of the peculiar--velocity field, then
the gravitational field equations are not needed to close the system
of equations (\ref{evolution}): the peculiar--velocity solves
\begin{equation}
\dot {\bf u} + (H-F(t)) {\bf u} = {\bf 0} \;\;\;.   
\label{zeldovich}
\end{equation}
If the field is appropriately scaled and a new time--variable is 
introduced (see below),
Zel'dovich's approximation manifests itself as an essientially force--free 
description of the continuum.

Gurbatov et al. (loc.cit.) proposed that one should add a forcing that is
directly proportional to the Laplacian of the peculiar--velocity field
to this equation which, in the scaled variables, attains the form of 
Burgers' equation. 
\smallskip\noindent
To derive the `adhesion approximation' from the effective kinetic equations
(\ref{evolution}) 
we just need to insert the law of motion (\ref{proportionality}) 
into the second of the equations (\ref{evolution}), 
which involves the multi--stream ``pressure'':

\begin{equation}
\dot {\bf u} + (H-F(t)) {\bf u} = \zeta F(t) {\bf \Delta}_{\bf q} {\bf u} 
\;\;\;;\;\;\;\zeta: = \frac{\alpha' (\varrho)}{a^2}\frac{1}{4\pi G\varrho}\;\;\;.   
\label{burgers1}
\end{equation}
In the `adhesion approximation' the function $F(t)$ is determined as in 
Zel'dovich's approximation by the requirement of reproducing the linear
solution of gravitational instability:

\begin{equation}
F(t) = 4\pi G\varrho_H {b(t)\over {\dot b}(t)} \;\;\;,
\label{parallelity}
\end{equation}
where $b(t)$ is identical to the growing density contrast 
mode solution of the Eulerian 
linear theory of gravitational instability for ``dust'' (i.e., it solves the equation
$\ddot{b} + 2 H \dot{b} - 4 \pi G \varrho_{H} b = 0$).

Changing the temporal variable from $t$ to $b$ and defining a
rescaled velocity field $\tilde{\bf u}: = {\bf u} / a
\dot{b}$, Equation (\ref{burgers1}) becomes the well--known key
equation of the `adhesion approximation' where $\mu$ is assumed constant 
(Gurbatov et al. loc.cit.):

\begin{equation}
\frac{d \tilde{\bf u}}{d b} = {\mu} {\bf \Delta}_{\bf q} \tilde{\bf u} \;\;\; , 
\;\;\;  {d \over db} := {\partial \over \partial b} + 
\tilde {\bf u} \cdot \nabla_{\bf q} \;\;\;,\;\;\;\mu := \frac{\zeta F(t)}{\dot b}
=\frac{\alpha' (\varrho)}{a^2}\frac{\varrho_H}{\varrho}\frac{b}{{\dot b}^2}\;\;\;.  
\label{burgers2}
\end{equation}

\subsection{The Lagrangian Linear Regime.}

The `adhesion approximation', as we saw above, can be derived 
from kinetic routes. Although this derivation is not a contrived experimenting
way of formally getting the Laplacian forcing, we may criticize it for its
limited range of validity in the kinetic framework. 
Buchert \& Dom\'\i nguez (loc.cit.)
have shown that the description has to be limited to small velocity dispersion
which is a necessary condition to still follow Zel'dovich's trajectories for the
bulk flow (the law of motion (\ref{proportionality})). Starting out from the 
full system of effective kinetic equations we may pursue a
systematic way of constructing models of `adhesion' by using the
Lagrangian perturbation theory. To the first order in the displacements from
a homogeneous--isotropic reference cosmology Adler \& Buchert (1999)
obtain for the longitudinal part of the (Lagrangian) displacement field
${\bf P}({\bf X},t)$: 
\begin{equation}
{\ddot{\bf P}} + 2 H {\dot{\bf P}} - 4\pi G \varrho_H {\bf P}= 
\frac{C_{\alpha}}{a^2} {\bf \Delta}_{\bf X}{\bf P}\;\;;\;\;C_{\alpha}: = \alpha' =
const.\;\;,\;\;{\dot{\bf X}} = {\bf 0}\;\;.
\label{pepsi}
\end{equation}
The markedly familiar differential operator in this equation 
helps to construct solutions of this Lagrangian linear equation 
from known solutions of the Eulerian linear theory.
The so constructed solutions may be employed as models for 
adhesive gravitational clustering in the weakly nonlinear regime. 

\subsection{New Adhesion.}

The apparent disparity between the standard `adhesion model' and the
Lagrangian perturbation approach can be made more transparent by 
reorganizing the general equations into a single equation for the 
gravitational peculiar--acceleration:
\begin{equation}
{\ddot{\bf w}} + 6 H {\dot{\bf w}} +(2 {\dot{H}} + 8 H^2 - 4\pi G \varrho_H )
{\bf w} = 4\pi G \varrho_H \zeta {\bf \Delta}_{\bf q} {\bf w}
+ \dot{\cal R} + 4H {\cal R} \;\;\;\;.
\label{beyondquasilinear1} 
\end{equation}
${\cal R}$ represents nonlinear residuals in the equations which are 
touched a little further in a paper in preparation (Buchert 1999).

Hence, we are led to suggesting the following model equation for 
adhesive gravitational clustering in the nonlinear regime, neglecting the
residuals:
\begin{equation}
{\ddot{\bf w}} + 6 H {\dot{\bf w}} +(2 {\dot{H}} + 8 H^2 - 4\pi G \varrho_H )
{\bf w} = 4\pi G \varrho_H \zeta {\bf \Delta}_{\bf q} {\bf w} \;\;\;.
\label{beyondquasilinear2} 
\end{equation}
A beautiful feature of this equation is that it embodies two limiting cases:
1. the standard `adhesion approximation' in the limit of small velocity 
dispersion, and 
2. the Lagrangian linear model (\ref{pepsi}) being a solution of 
\begin{equation}
{\ddot{\bf w}} + 6 H {\dot{\bf w}} +(2 {\dot{H}} + 8 H^2 - 4\pi G \varrho_H )
{\bf w} = \frac{C_{\alpha}}{a^2} {\bf \Delta}_{\bf X} {\bf w}\;\;\;.
\label{beyondquasilinear3} 
\end{equation}
The proposed new models will have, by construction,
an improved performance for the modeling of large--scale structure, but 
they will hopefully also give more insight into the clustering properties at
the stages of 
stabilization of large--scale structure, previrialization and the onset to 
``virialized'' systems. A further clue might be inferred with regard to 
the possible emergence of N--soliton states (compare G\"otz (1988) for a
one--dimensional example).  

\newpage

\acknowledgments

It is a pleasure to thank the organizers of this workshop for a pleasant stay 
at Hope Hotel, Shanghai, and for guided sightseeing. 
Thanks to Tom Abel, Thomas Boller, Karsten Jedamzik and Ravi Sheth for
``deepening'' discussions and worth--to--remember Shanghai tours. 
The Max--Planck--Society and the Chinese Academy
of Sciences are acknowledged for their financial
support. This work is part of a project in the 
``Sonderforschungsbereich f\"ur Astroteilchenphysik''  SFB 375, Munich.

\end{document}